\documentclass[preprint, trackchanges]{aastex702} 

\usepackage{multirow, url, color}
\usepackage{bbding, amssymb} 
\usepackage{ulem} 
\usepackage{verbatim} 
\usepackage{enumerate}
\usepackage{amsmath} 
\usepackage{booktabs} 
\usepackage{graphicx, subfigure}
\usepackage{tikz}
\usetikzlibrary{arrows.meta}
\usepackage[fleqn]{mathtools}
\usepackage{mathrsfs}
\usepackage[makeroom]{cancel}

\newcommand{\di}{\mathrm{d}}
\newcommand{\bs}[1]{\boldsymbol{#1}}
\newcommand{\dt}{\Delta t}
\newcommand{\dH}{\delta\widetilde{H}}
\newcommand{\pH}{\widetilde{H}}
\shorttitle{Implicit Poisson Bracket Scheme for Steady-State GCRs}
\shortauthors{Liu \& Sokolov}

\begin{document}

\title{An Implicit Conservative Finite Volume Scheme with Poisson Brackets for the Steady-State Transport of Galactic Cosmic Rays}

\correspondingauthor{Weihao Liu}
\email[show]{whliu@umich.edu}

\author[orcid=0000-0002-2873-5688]{Weihao Liu}
\affiliation{Department of Climate and Space Sciences and Engineering, University of Michigan, Ann Arbor, MI 48109, USA}
\email{whliu@umich.edu}
\author[orcid=0000-0002-6118-0469]{Igor V. Sokolov}
\affiliation{Department of Climate and Space Sciences and Engineering, University of Michigan, Ann Arbor, MI 48109, USA}
\email{igorsok@umich.edu}

\begin{abstract}
Galactic cosmic rays (GCRs) pose a persistent radiation hazard to astronauts and spacecraft in the heliosphere. Modeling their transport is essential for understanding how the solar wind and heliospheric magnetic field shape their spatial distributions and energy spectra. Timely predictions require solving for the particle distribution function much faster than real time, but explicit numerical schemes are restricted by the stability limit on the time step. 
In this Letter, we present an implicit conservative finite-volume scheme for kinetic equations formulated with Poisson brackets. 
This scheme conserves particle number, maintains the total-variation-diminishing property, achieves second-order spatial accuracy, and remains stable for time steps far beyond the explicit limit, while reaching quasi-steady states in $\lesssim\! 1/100$ of the computational time required for the explicit scheme, as demonstrated by numerical tests. 
We apply this scheme to the Parker transport equation along magnetic field lines in the background solar wind during Carrington rotation $2211$, where it reaches steady-state GCR distributions within $10$ iterations, about $1000$ times faster than the explicit scheme. The GCR proton spectrum averaged over $27$ days agrees with AMS-02 measurements to within $10\%$. The simulation also yields a realistic radial intensity gradient and captures the Forbush decrease features associated with three stream interaction regions observed at Earth. 
\end{abstract}


\section{Introduction} \label{sec1:intro}

In interplanetary space, galactic cosmic rays (GCRs) and solar energetic particles (SEPs) are the main sources of particle radiation and can pose hazards to astronauts and spacecraft \citep{cucinotta2006cancer, vainio2009dynamics}. GCRs originate from galactic sources such as supernova remnants \citep{ackermann2013detection} and provide a persistent radiation background modulated by transport through the heliosphere, whereas SEPs arise from solar flares and coronal mass ejection (CME)-driven shocks, producing transient radiation level enhancements \citep{desai2016large, reames2021solar}. Large SEP events are produced mainly by diffusive shock acceleration \citep[DSA;][]{fermi1949origin, krymskii1977regular, axford1977acceleration, bell1978acceleration1, blandford1978particle}. Accurate and efficient modeling of particle acceleration and transport is essential for assessing these radiation hazards. 

The acceleration and transport of these energetic particles are described by kinetic equations for their phase-space distribution function. 
In the diffusive limit, the Parker transport equation \citep{parker1965passage} governs the omnidirectional distribution through convection, spatial diffusion, and adiabatic momentum changes, accounting for DSA at shocks and cooling in the expanding solar wind \citep{drury1983introduction}. 
In numerical modeling, particle number conservation and computational efficiency are important. The convective and adiabatic terms can be written using Poisson brackets (PBs), which provide a formulation of the Liouville equation that keeps the distribution function constant along particle trajectories and conserves particle number for Hamiltonian motion \citep{liouville1838note, landau1959course}. Combining this formulation with finite volume methods \citep[FVMs;][]{hirsch2007numerical} allows the conservation law to be preserved in the numerical discretization. 

Using integral relations for PBs, \cite{sokolov2023high} developed an explicit, conservative FVM scheme that conserves the particle number and maintains the total-variation-diminishing \citep[TVD;][]{toth2023total} property. 
\cite{liu2025physics} further implemented this PB scheme in the Multiple Field-Line Advection Model for Particle Acceleration \citep[M-FLAMPA;][]{sokolov2004new, borovikov2018toward} within the Space Weather Modeling Framework \citep[SWMF;][]{toth2005space, toth2012adaptive} to simulate SEP acceleration and transport in the evolving corona and heliosphere \citep[see also][]{liu2026counterintuitive}. M-FLAMPA also serves as the SEP component of the Solar wind with FIeld lines and Energetic particles (SOFIE) model, which has been applied to historical SEP events \citep{zhao2024solar} and tested in a simulated operational setting for SEP prediction \citep{liu2026testbed}. However, the explicit scheme is subject to the Courant--Friedrichs--Lewy (CFL) condition \citep{courant1928partiellen}. This restriction becomes severe where phase-space advection rates are large relative to cell sizes, such as at shocks and near the Sun. Consequently, simulations can require many small time steps to accelerate particles to high energies or to reach a steady-state distribution, limiting their computational efficiency. 

GCR transport is one such steady-state problem \citep{kota2013theory, engelbrecht2022theory}. GCRs consist mostly of protons, are believed to originate mainly in supernova remnants, and enter the heliosphere from the local interstellar medium \citep{blasi2013origin}. Within the heliosphere, they undergo solar wind convection, diffusion, magnetic drifts, and adiabatic energy losses. Together, these processes produce solar modulation, which generally lowers GCR intensities relative to their interstellar values and varies with solar activity \citep{potgieter2013solar, wang2022variation, wang2025rigidity}. 
Modulation models range from empirical approaches \citep[e.g.,][]{matthia2013ready, gieseler2017} and the force-field approximation \citep[e.g.,][]{gleeson1968solar, corti2016, shen2021solar} to numerical solutions of the Parker transport equation using finite differences \citep[e.g.,][]{fisk1971solar, kota1983effects, badhwar2015, pelosi2026new} or stochastic differential equations \citep[e.g.,][]{zhang1999markov, pei2010general, strauss2012modelling, shen2018modulation, boschini2022transport, alonsoguzman2024numerical}, which are evaluated against particle spectra and radiation measurements \citep[e.g.,][]{corti2019numerical, whitman2019comparison, liu2023modeling, liu2024comprehensive, vaisanen2025simulation}.

Beyond the spectrum at Earth, temporal variations and spatial gradients provide further constraints on GCR modulation. 
The heliospheric magnetic field evolves with solar activity and reverses its large-scale polarity approximately every 11 years \citep{babcock1959sun, cane1999cosmic, ferreira2004long}, and GCR intensities generally anticorrelate with solar activity \citep{forbush1954world, forbush1958cosmic}. 
On timescales of days, CMEs and stream interaction regions (SIRs) can produce temporary reductions in GCR intensity, known as Forbush decreases \citep[FDs;][]{forbush1937, lockwood1971forbush}, through changes in particle convection and diffusion associated with enhanced magnetic fields and turbulence \citep{cane2000fd, richardson2004energetic, richardson2011, luo2017numerical, luo2020numerical, alonsoguzman2026modulation}. 
GCR intensities also generally increase with heliocentric distance and vary with latitude, with gradients that depend on particle rigidity, solar activity, and magnetic polarity \citep[e.g.,][]{gieseler2016spatial, honig2019multi, Roussos2020, xu2026anomalous}. 
Near the Sun, GCRs produce gamma-ray disk emission through hadronic interactions with the solar atmosphere, while GCR electrons produce an extended component through inverse-Compton scattering of sunlight \citep{abdo2011fermi}. The observed disk emission exhibits features in its intensity, spectrum, and solar-cycle variation that are incompletely understood \citep[e.g.,][]{ng2016first, linden2018evidence, tang2018unexpected}. 
Except for FDs caused by CMEs, these signals vary slowly with solar rotation, because the ambient solar wind is nearly steady in the frame corotating with the Sun. Modeling them therefore requires the steady-state GCR distribution throughout the inner heliosphere down to the low corona, but the current explicit PB scheme converges slowly and requires $\gtrsim\!10^4$ time steps to reach this steady state.

In this Letter, we develop an implicit conservative PB scheme that remains stable for time steps far beyond the CFL limit, test it on relativistic proton gyration and on particle acceleration at an analytical shock, and apply it to steady-state GCR transport. Section~\ref{sec2:scheme} describes the scheme and summarizes its tests, which are detailed in Appendix~\ref{app:tests}. Section~\ref{sec3:gcr} applies the scheme to GCR protons in the steady-state solar wind of Carrington rotation (CR) $2211$ and compares the results with observations at Earth. Section~\ref{sec4:summary} summarizes the main findings and discusses their implications for modeling GCRs and SEPs.

\section{The Implicit Poisson Bracket Scheme} \label{sec2:scheme}

    \subsection{Conservative Fluxes and the Implicit Step} \label{sec:fv}

    For a system with $L$ degrees of freedom and canonical coordinate–momentum pairs $(q_\ell,p_\ell), \, \ell = 1, \dots, L$, the Liouville equation reads
    \begin{equation}
        \frac{\partial f}{\partial t} + \sum_{\ell=1}^{L}\{f;H\}_{q_\ell,\, p_\ell} = 0, \quad
        \{f;H\}_{q_\ell,\, p_\ell} = \frac{\partial f}{\partial q_\ell}\frac{\partial H}{\partial p_\ell} - \frac{\partial f}{\partial p_\ell}\frac{\partial H}{\partial q_\ell}, \label{eqn:liouville}
    \end{equation}
    where $H$ is the Hamiltonian. By Stokes' theorem, the integral of each PB over a phase-space cell (\textit{control volume}) $i$ of volume $V_i$ equals the line integral of $f\,\di H$ around its boundary in the $(q_\ell,\, p_\ell)$ plane \citep{sokolov2023high}. In this plane, each side of cell $i$ forms a shared face with a neighboring cell $j$. Using a reconstructed value $f^{\rm face}_{ij}$ on the shared face, the particle number $N_i = V_i f_i$ evolves as
    \begin{equation}
        \frac{\di N_i}{\di t}=-\sum_{j\in\mathbb{N}_i}\Phi_{ij},\quad
        \Phi_{ij}=\dH_{ij}\,f^{\rm face}_{ij}, \label{eqn:balance}
    \end{equation}
    where $\mathbb{N}_i$ is the set of neighbors of cell $i$, $\dH_{ij}$ denotes the increment of the effective Hamiltonian $\pH_\ell$ along the corresponding segment, and $\pH_\ell$ is obtained by integrating $H$ over the remaining phase-space coordinates. Since fluxes across shared faces enter adjacent cells with opposite signs, internal exchanges cancel and the total particle number is conserved in the absence of net boundary fluxes, as in finite volume methods for conservation laws \citep{hirsch2007numerical}.

    Particles leave cell $i$ through faces with positive fluxes, $\dH_{ij} > 0$. We consider three numerical fluxes here. The upwind flux takes $f_i$ on these outflow faces and $f_j$ on the inflow faces, giving the first-order operator, $\di f_i/\di t = -\mathcal{L}^{(1)}_{\rm PB}f_i$, with the local CFL number $C_i=\dt A_i/V_i$, where $\dt$ denotes the time step and $A_i$ is the sum of $\dH_{ij}$ over the outflow faces. The central-difference flux takes the face value $\left(f_i+f_j\right)/2$, giving a second-order operator $\mathcal{L}^{(2)}_{\rm PB}$, which can produce spurious oscillations near steep gradients. The TVD flux adds to the upwind flux an anti-diffusive correction, $\mathcal{L}^{(\rm lim2)}_{\rm PB}$, limited to avoid these oscillations. We treat the upwind operator implicitly and the correction explicitly, which avoids solving a highly nonlinear system for the flux-limiter correction. With a time weight $0\leqslant \alpha \leqslant1$ that weights the upwind operator between the old ($\bs{f}^{n}$) and new ($\bs{f}^{n+1}$) time levels, the update reads
    \begin{equation}
        \left( \mathcal{I} + \alpha\dt\,\mathcal{L}^{(1)}_{\rm PB} \right) \bs{f}^{n+1}
        = \left[ \mathcal{I} - (1-\alpha)\dt\,\mathcal{L}^{(1)}_{\rm PB} \right] \bs{f}^{n}
        + \dt\,\mathcal{L}^{(\rm lim2)}_{\rm PB}\bs{f}^{n}, \label{eqn:update}
    \end{equation}
    where $\mathcal{I}$ is the identity operator, $\bs{f}^{n}$ collects values of $f$ in all cells at time level $n$, and $\alpha=0$, $1/2$, and $1$ give three time-stepping methods, the fully explicit (forward Euler), Crank--Nicolson, and fully implicit (backward Euler) schemes, respectively. In $\mathcal{L}^{(\rm lim2)}_{\rm PB}$, we apply a $\beta$-limiter \citep{van1977towards} to the anti-diffusive fluxes and the Koren limiter \citep{koren1993robust} to the higher-order reconstruction to prevent new extrema and maintain the TVD property \citep[see][Section 2.5]{sokolov2023high}. 

    For any $\dt>0$, the matrix on the left-hand side in Equation~\eqref{eqn:update} is sparse and strictly diagonally dominant, with positive diagonal and nonpositive off-diagonal entries. It is therefore a nonsingular M-matrix with a nonnegative inverse \citep{berman1994nonnegative}, and the first-order update preserves $f \geqslant 0$ and remains stable when 
    \begin{align}
        (1-\alpha)C_i \leqslant 1,\, \forall i. \label{eqn:cfl}
    \end{align}
    For $\alpha<1$, Equation~\eqref{eqn:cfl} requires $C_i \leqslant 1/(1-\alpha)$, giving limited time steps, whereas the fully implicit scheme ($\alpha=1$) allows sufficiently large time steps for steady-state iterations without this restriction. 
    For $\alpha > 0$, we solve the sparse linear system in Equation~\eqref{eqn:update} at each time step using Gauss--Seidel iterations, which converge since the coefficient matrix is diagonally dominant. The update is generally first-order accurate in time. For $\alpha = 1/2$, we perform additional Picard iterations to time-center the correction and recover second-order temporal accuracy. 
    The scheme performance is then verified by a test of relativistic proton gyration, as detailed in Appendix~\ref{app:gyration}. Hereafter, we refer to the TVD flux with forward and backward Euler time stepping as the explicit and implicit PB schemes, respectively. 

    \subsection{The Parker Transport Equation along a Magnetic Field Line} \label{sec:parker}

    The Parker transport equation \citep{parker1965passage} describes the distribution of energetic particles in the diffusive limit, where their distribution is omnidirectional. Neglecting perpendicular diffusion and drifts, we solve for the distribution function, $f(t,s,p)$, of particles with momentum $p$ at time $t$ and arc length $s$ along individual magnetic field lines:
    \begin{equation}
        \frac{\partial f}{\partial t} + B\left\{f;\,\frac{\rho u}{B}\,\frac{p^3}{3\rho}\right\}_{s,\,p^3/3}
        = B\,\frac{\partial}{\partial s}\left(\frac{\kappa_\parallel}{B}\,\frac{\partial f}{\partial s}\right).
        \label{eqn:parker}
    \end{equation}
    Here, $u$, $\rho$, and $B$ denote the plasma speed, density, and magnetic field strength, respectively. $\kappa_\parallel = v\lambda_\parallel/3$ is the parallel diffusion coefficient, with $v$ the particle speed and $\lambda_\parallel$ the parallel mean free path. 
    The PB on the left-hand side combines convection and adiabatic energy change, while the right-hand side describes parallel diffusion. We test the PB schemes for Equation~\eqref{eqn:parker} by simulating DSA at a planar shock (Appendix~\ref{app:shock}). Both the explicit and implicit PB schemes reproduce the analytical steady-state spectrum, while the implicit scheme reaches the steady state $\sim\!50$ and more than $700$ times faster than the explicit scheme at $C=98$ and $9800$, respectively. 

    For a steady-state, stream-aligned solar wind background in the frame corotating with the Sun, where the plasma velocity is parallel or antiparallel to the magnetic field \citep{parker1958dynamics, sokolov2022stream}, we march Equation~\eqref{eqn:parker} to its steady state by advancing the first-order PB and diffusion operators together implicitly, while evaluating the limited anti-diffusive correction at the previous time step, 
    \begin{equation}
        \left[\mathcal{I}+\dt\left(\mathcal{L}^{(1)}_{\rm PB}+\mathcal{L}_{\rm D}\right)\right]\bs{f}^{n+1}
        = \bs{f}^{n}+\dt\,\mathcal{L}^{(\rm lim2)}_{\rm PB}\bs{f}^{n},
        \label{eqn:monolithic}
    \end{equation}
    where $-\mathcal{L}_{\rm D}\bs{f}$ is the three-point discretization of the diffusion term. The matrix on the left-hand side of Equation~\eqref{eqn:monolithic} remains an M-matrix, and the fixed point of this update satisfies
    \begin{equation}
        \left(\mathcal{L}^{(1)}_{\rm PB}+\mathcal{L}_{\rm D} - \mathcal{L}^{(\rm lim2)}_{\rm PB}\right)\bs{f} = \bs{0}, \label{eqn:fixedpoint}
    \end{equation}
    which is independent of $\dt$, so we use local time stepping $\dt_i=CV_i/A_i$ for each cell with one CFL number $C$ when solving for the steady-state GCR distribution. 

\section{GCR Propagation Through the Background Solar Wind} \label{sec3:gcr}

    \subsection{Simulation Setup} \label{sec:gcr_setup}

    To model GCRs in a realistic heliosphere, we use a steady, stream-aligned solar wind solution \citep{sokolov2022stream} for CR~$2211$ ($2018$ November $22$--December $19$), computed within the SWMF using the Alfv\'en Wave Solar atmosphere Model-Realtime \citep[AWSoM-R;][]{sokolov2013magnetohydrodynamic, sokolov2021threaded, van2014alfven} from $1.1$ solar radii in the low corona to $5\; \mathrm{au}$ in the inner heliosphere \citep[see][for the adopted grid setup]{shou2017new}. 
    The resulting solar wind parameters and magnetic field lines are coupled to M-FLAMPA \citep{sokolov2004new, liu2025physics}, which solves Equation~\eqref{eqn:parker} along each field line. Following \cite{li2003energetic} and \cite{borovikov2019toward}, we prescribe the parallel mean free path as 
    \begin{equation}
        \lambda_\parallel=\lambda_0 \left(\frac{r}{1~\mathrm{au}}\right)\left(\frac{pc}{1~\mathrm{GeV}}\right)^{1/3}, \label{eqn:mfp}
    \end{equation}
    where $r$ is the heliocentric distance, and we adopt $\lambda_0 = 0.3\; \mathrm{au}$ for the reference mean free path parameter, consistent with \cite{chen2024parallel} and \cite{liu2025physics}. The momentum grid contains 100 logarithmically spaced bins with kinetic energies from $1~$MeV to $100~$TeV. 

    At the outer boundary, we prescribe the differential intensity $J=p^2f$ of GCR protons by applying force-field modulation \citep{gleeson1968solar} to the local interstellar spectrum of \citet{burger2000rigidity}, as parameterized by \citet{usoskin2005}. We adopt a modulation potential of $295\; \mathrm{MV}$, determined by minimizing the reduced $\chi^2$ between the GCR proton spectra at $1\; \mathrm{au}$ modeled by M-FLAMPA and those measured by the Alpha Magnetic Spectrometer \citep[AMS-02;][]{aguilar2021periodicities} averaged over CR~$2211$. We impose a zero-gradient condition at the inner boundary and extrapolate $f$ as a power law at both momentum boundaries. 
    We run the implicit PB scheme at $C=9800$ for $10$ iterations and, for comparison, the explicit scheme at $C=0.98$ for $30{,}000$ iterations. Both runs reach quasi-steady states with $L_1$-norm iterative residuals below $10^{-10}$, and the implicit run requires only $\sim\! 0.1\%$ of the CPU time of the explicit run.

    \subsection{GCR Proton Spectrum and Radial Gradient} \label{sec:gcr_result}

\begin{figure*}[tp!]
    \centering
    \includegraphics[width=0.975\hsize]{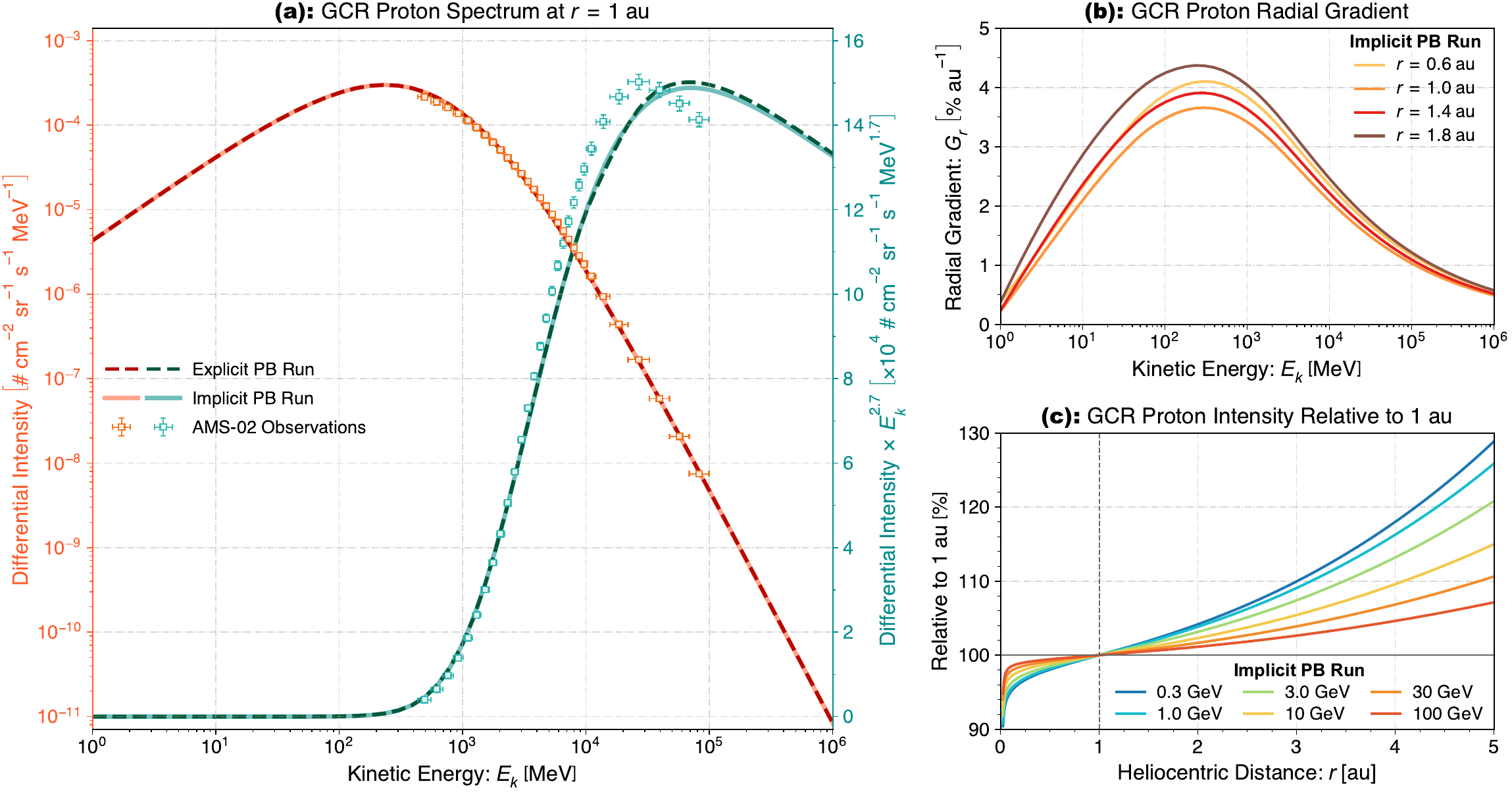}
    \caption{GCR proton distribution averaged along Earth's trajectory during CR~$2211$. 
    (a) Differential intensity at $1\; \mathrm{au}$, shown directly (left axis; red) and multiplied by $E_\mathrm{k}^{2.7}$ (right axis; green), with the latter highlighting differences at high energies and $E_\mathrm{k}$ denoting the kinetic energy. The lighter solid and darker dashed curves show the implicit solution after $10$ iterations at $C=9800$ and the explicit solution after $30{,}000$ iterations at $C=0.98$, respectively. The scattered points show AMS-02 measurements averaged over CR~$2211$.
    (b) Radial gradient ($G_r$) of the modeled GCR proton intensity at four representative heliocentric distances. 
    (c) Intensity at six selected energies expressed as a percentage of its value at $1\; \mathrm{au}$. Panels~(b) and (c) show results from the implicit PB scheme.} \label{fig1:gcr}
    \end{figure*}

    We extract GCR proton spectra along Earth's trajectory and average them over CR~$2211$ for comparison with AMS-02. As shown in Figure~\ref{fig1:gcr}(a), the spectra from explicit and implicit PB schemes agree within $0.05\%$ from $10~$MeV to $10~$GeV and within $2\%$ at higher energies. They both match the AMS-02 measurements within $10\%$, with some noticeable discrepancies at higher energies that may reflect limitations of the force-field boundary approximation \citep{moraal2013}, the adopted rigidity dependence of $\lambda_\parallel$, and the omission of perpendicular diffusion and particle drifts. Hereafter, we present only the implicit solution.

    Averaged over CR~$2211$, the modeled radial gradient $G_r = \frac{1}{J} \frac{\partial J}{\partial r} = \frac{\partial\ln J}{\partial r}$ ranges from $\sim\!0.5$ to $4.5\%\; \mathrm{au}^{-1}$ across all four heliocentric distances in the equatorial plane, peaking near $300$~MeV (rigidity $\simeq0.8$~GV), as shown in Figure~\ref{fig1:gcr}(b).
    The peak values, $3.5$--$4.5\%\; \mathrm{au}^{-1}$, are comparable to the $3.0$--$4.5\%~\mathrm{au}^{-1}$ GCR gradients reported in previous studies \citep{vos2016global, honig2019multi, Roussos2020}, despite differences in spatial and temporal coverage. 
    
    Figure~\ref{fig1:gcr}(c) shows the radial profiles of GCR proton intensity, normalized to their values at $1\; \mathrm{au}$. Between $1$ and $5\; \mathrm{au}$, the intensity increases by $\sim\! 28\%$ at $0.3~$GeV and $7\%$ at $100~$GeV, reflecting stronger solar modulation for lower-energy particles.
    For a given energy, the intensity profile in panel~(c) is steep close to the Sun, flattens near $1\; \mathrm{au}$, and steepens again farther out. The local slope divided by the normalized intensity gives $G_r$, which reaches a minimum near $r = 1\; \mathrm{au}$. This radial variation is also reflected in the overall level of the gradient curves in panel~(b), where the $1\; \mathrm{au}$ curve lies below the other curves. 

    These trends can be interpreted within the force-field approximation \citep{gleeson1968solar}. For an idealized equatorial Parker spiral with radial solar wind speed ($U_r$) and diffusion only along the magnetic field, 
    \begin{equation}
        G_r\approx - \frac{U_{r}}{3\kappa_{rr}} \frac{\partial\ln f}{\partial\ln p},
        \quad
        \kappa_{rr}=\kappa_\parallel\cos^2\psi = \frac{v\lambda_\parallel/3}{1+(\Omega r/U_{r})^2}, \label{eqn:gradient}
    \end{equation}
    where $\kappa_{rr}$ is the radial diffusion coefficient \citep{jokipii1971propagation}, $\psi$ is the magnetic field angle relative to the radial direction \citep{parker1958dynamics}, and $\Omega = 2.87\times10^{-6}\; \mathrm{s}^{-1}$ denotes the angular frequency of solar rotation. 
    At high particle energies, where $v\approx c$ and the spectral slope varies weakly, the prescribed $\lambda_\parallel\propto p^{1/3}$ gives $G_r\propto p^{-1/3}$. More efficient diffusion therefore explains the weaker radial variation at higher energies in panel~(c). 

    Besides, the prescribed $\lambda_\parallel\propto r$ (Equation~\eqref{eqn:mfp}) and the Parker spiral geometry help explain the modeled local minimum in $G_r$ near $1\; \mathrm{au}$. Closer to the Sun, the shorter mean free path reduces diffusion; farther out, the increasingly azimuthal magnetic field reduces its radial component (Equation~\eqref{eqn:gradient}). These effects produce small $\kappa_{rr}$ at small or large heliocentric distance, with a maximum in between. For a constant $U_{r} = 400\; \mathrm{km\,s}^{-1}$, this maximum occurs at $r=U_{r}/\Omega \simeq 0.93\; \mathrm{au}$. If the logarithmic spectral slope, $\frac{\partial\ln f}{\partial\ln p}$, varies weakly with radius, Equation~\eqref{eqn:gradient} gives a corresponding local minimum in $G_r$, consistent with panel~(b). 
    Perpendicular diffusion, omitted here, would add $\kappa_\perp\sin^2\psi$ to $\kappa_{rr}$ and could reduce the larger radial gradients farther from the Sun. In future work, we will implement perpendicular diffusion and assess its effects on GCR modulation and spatial gradients.

    \subsection{GCR Modulation over CR 2211} \label{sec:gcr_map}

    \begin{figure*}[tp!]
    \centering
    \includegraphics[width=0.975\hsize]{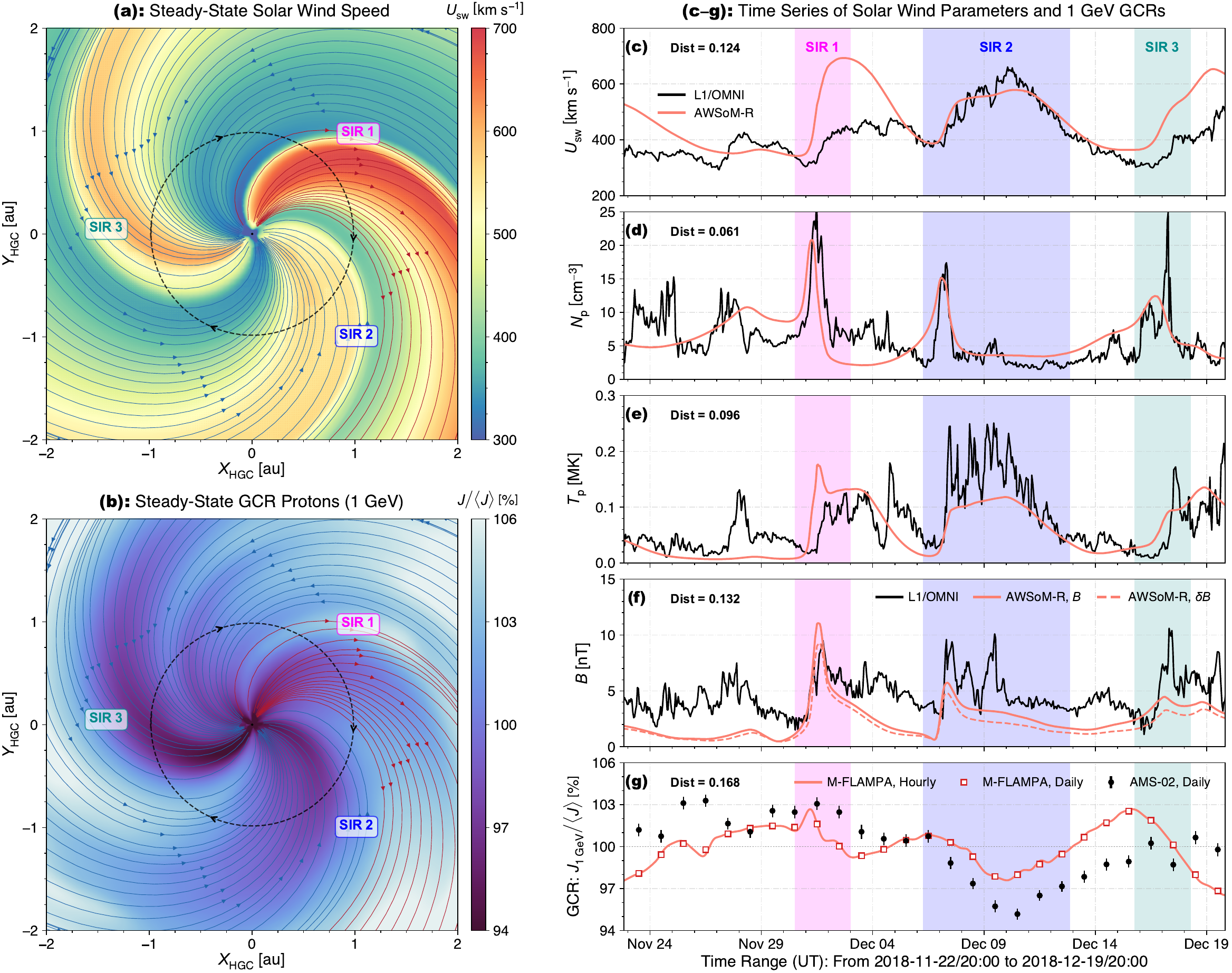}
    \caption{GCR simulation results in the steady-state solar wind during CR~$2211$. 
    (a) Solar wind speed ($U_\mathrm{sw}$) in the solar equatorial plane modeled by AWSoM-R, with magnetic field lines colored by their polarity and Earth's trajectory marked by the dashed curve.
    (b) GCR proton intensity at $1~$GeV modeled by M-FLAMPA, normalized to its average along Earth's trajectory over CR~$2211$ ($J/\langle J\rangle$). Panels (a) and (b) are plotted in the heliographic Carrington (HGC) coordinates, which corotate with the Sun.
    (c)--(g) Time profiles at Earth of the solar wind speed ($U_\mathrm{sw}$), proton density ($N_\mathrm{p}$), proton temperature ($T_\mathrm{p}$), magnetic field strength ($B$), and $1~$GeV GCR proton intensity relative to its average at Earth, marked with the distance values between the simulation results (red) and observational data (black) defined by \cite{sachdeva2019validation}. 
    In panel (f), the red solid curve shows the total modeled magnetic field strength including fluctuation ($B$), while the dashed curve shows the contribution from Alfv\'en wave turbulence ($\delta B$).
    In all panels, three SIRs are marked by pink, purple, and green labels and/or shaded regions, respectively.} \label{fig2:map}
    \end{figure*}

    We next examine GCR intensity variations during CR~$2211$ by interpolating the steady-state solution along Earth's trajectory in the corotating frame. Figure~\ref{fig2:map} shows the solar wind speed and $1~$GeV GCR proton intensity in the equatorial plane, together with time series at Earth. The AWSoM-R solution captures three SIRs identified in the L1/OMNI data \citep{king2005solar}, although the solar wind speed is overestimated during SIR~$1$ (panel~(c)).
    We quantify the model--observation agreement using the distance metric in \citet{sachdeva2019validation}, obtaining values of $0.061$--$0.168$ for the five profiles in panels~(c)--(g). These distance values remain relatively small ($\lesssim 0.15$) and comparable to those reported in previous ambient solar wind simulations \citep[e.g.,][]{sachdeva2023solar, huang2024solar, liu2026counterintuitive}. Together with the reproduced SIR structures, they suggest reasonable overall agreement with the observations at Earth.

    Within the modeled inner heliosphere, the GCR intensity generally increases with heliocentric distance (panel~(b)). Along Earth's trajectory, the $1~$GeV GCR proton intensity varies by about $\pm3\%$ around its mean. As Earth passes through the three SIRs, the modeled intensity decreases by $2\%$--$6\%$, producing FD features comparable to those in the daily AMS-02 measurements (panel~(g)). These decreases reflect GCR modulation by the compressed plasma and magnetic fields in SIRs \citep{richardson2004energetic, guo2014corotating, luo2020numerical, vrsnak2022analytic}.

\section{Summary and Discussion} \label{sec4:summary}

We develop an implicit, conservative FVM scheme for kinetic equations formulated with PBs. As verified by numerical tests, the proposed scheme conserves particle number, achieves second-order spatial accuracy, maintains the TVD property, and remains stable for time steps far beyond the CFL limit, with one sparse linear solve by Gauss--Seidel iterations per time step, requiring CPU times of only $\lesssim\! 1/100$ of those used by the explicit scheme. 

In M-FLAMPA, we implement this scheme to solve the Parker transport equation for GCR protons in the background solar wind of CR~$2211$. At a large CFL number ($C=9800$), the scheme reaches a quasi-steady distribution within $10$ iterations, requiring only seconds of CPU time per field line. Averaged along Earth's trajectory over CR~$2211$, the modeled spectrum at $1\;\mathrm{au}$ agrees with AMS-02 measurements within $\sim\!10\%$ above $300~$MeV. The simulation also yields radial gradients comparable in magnitude to earlier observational estimates and captures FD features due to the modulation by three SIRs, with amplitudes comparable to the observations. 

Note that our present GCR simulations neglect perpendicular diffusion \citep{jokipii1966cosmic, matthaeus2003nonlinear, shalchi2020perpendicular}, which could reduce the larger radial gradients beyond $\sim\!1$~au, and particle drifts \citep{jokipii1977effects, potgieter1985drift, troskie2024cosmic}, which introduce dependence on particle charge sign and solar magnetic polarity into GCR modulation. 
Additional uncertainties arise from the prescribed parallel mean free path and the force-field representation of the outer boundary at $5\; \mathrm{au}$. Future work will incorporate these omitted processes and assess their influence on the modeled spectra and gradients. 

The computational efficiency of the implicit PB scheme benefits the prediction of energetic particle fluxes. 
In applications, given an available solar wind solution \citep[e.g.,][]{owens2022rate, huang2024solar, koban2025dailyAGU}, real-time updates of steady-state GCR distributions become feasible. These solutions can provide the GCR component of the background seed particle population in SEP acceleration models, supporting efforts to improve SEP forecasting \citep[e.g.,][]{desai2006seed, hu2017modeling, wijsen2023seed, georgoulis2024prediction, chen2025evidence, liu2026testbed}. Simulations over successive Carrington rotations can also support studies of long-term GCR modulation and provide inputs for space radiation assessments. 

\begin{acknowledgments}
This work is supported by the NSF ANSWERS grant GEO-2149771 and the NASA Space Weather Center of Excellence program under award 80NSSC23M0191 (CLEAR). The authors thank the AMS-02 team for providing valuable data. The authors also thank Dr. Claudio Corti (University of Hawaii at Manoa; NASA Goddard Space Flight Center), Dr. Jamie S. Rankin (Princeton University), and Prof. Ofer Cohen (University of Massachusetts Lowell) for helpful discussions. 
Computational resources supporting this work are provided by the NASA High-End Computing (HEC) Program through the NASA Advanced Supercomputing (NAS) Division at the Ames Research Center. 
\end{acknowledgments}

\appendix
\restartappendixnumbering

\section{Numerical Tests} \label{app:tests}

This appendix presents two tests of the proposed scheme. 
First, we perform a test of relativistic proton gyration, using the exact solution to assess the accuracy, particle number conservation, and TVD property of the three numerical fluxes combined with the three time-stepping methods described in Section~\ref{sec:fv} (Appendix~\ref{app:gyration}).
Second, we perform a test of DSA at a planar shock, using the analytical steady-state spectrum to assess the accuracy and computational efficiency of the implicit PB scheme described in Section~\ref{sec:parker} at large time steps (Appendix~\ref{app:shock}).

    \subsection{Relativistic Proton Gyration} \label{app:gyration}
    
    We consider relativistic proton gyration in a uniform magnetic field, which points along $\hat{z}$. The protons move in the $(x, y)$ plane and gyrate clockwise when viewed from the positive $z$-axis toward the origin. 
    The proton charge ($e$) and rest mass ($m_0$), magnetic field strength ($B$), and speed of light ($c$) are normalized to unity, and time is measured in units of the inverse nonrelativistic gyrofrequency $\Omega_{\rm c}^{-1}$. 
    In the Landau gauge and for zero canonical momentum along $y$, the dynamics reduce to one canonical pair, $q=x$ and $p=p_x$, with Hamiltonian $H = \sqrt{1+q^2+p^2}$ \citep[][Section 16]{landau1975}. 
    The canonical equations, $\di q/\di t=p/H$ and $\di p/\di t=-q/H$, describe clockwise circular trajectories in the $(q,p)$ plane, giving the exact solution
    \begin{equation}
        f(t,q,p) = f\left(0,\,p\cos(\omega t)+q\sin(\omega t),\,q\cos(\omega t)-p\sin(\omega t)\right), \label{eqn:fexact}
    \end{equation}
    where the angular frequency $\omega = (1+q^2+p^2)^{-1/2} = 1/H$ is constant along each trajectory. 
    We initialize a smooth cap centered at $(q, p) = (-6,0)$ on the domain $[-12,12]^2$,
    \begin{equation}
        f_0=
        \begin{cases}
            \cos^4[\pi(q+6)/6] \cos^4(\pi p/6), & |q+6| \leqslant 3,\ |p| \leqslant 3,\\
            0, & \text{otherwise}.
        \end{cases} \label{eqn:finit}
    \end{equation}

    \begin{figure*}[tp!]
    \centering
    \includegraphics[width=0.975\hsize]{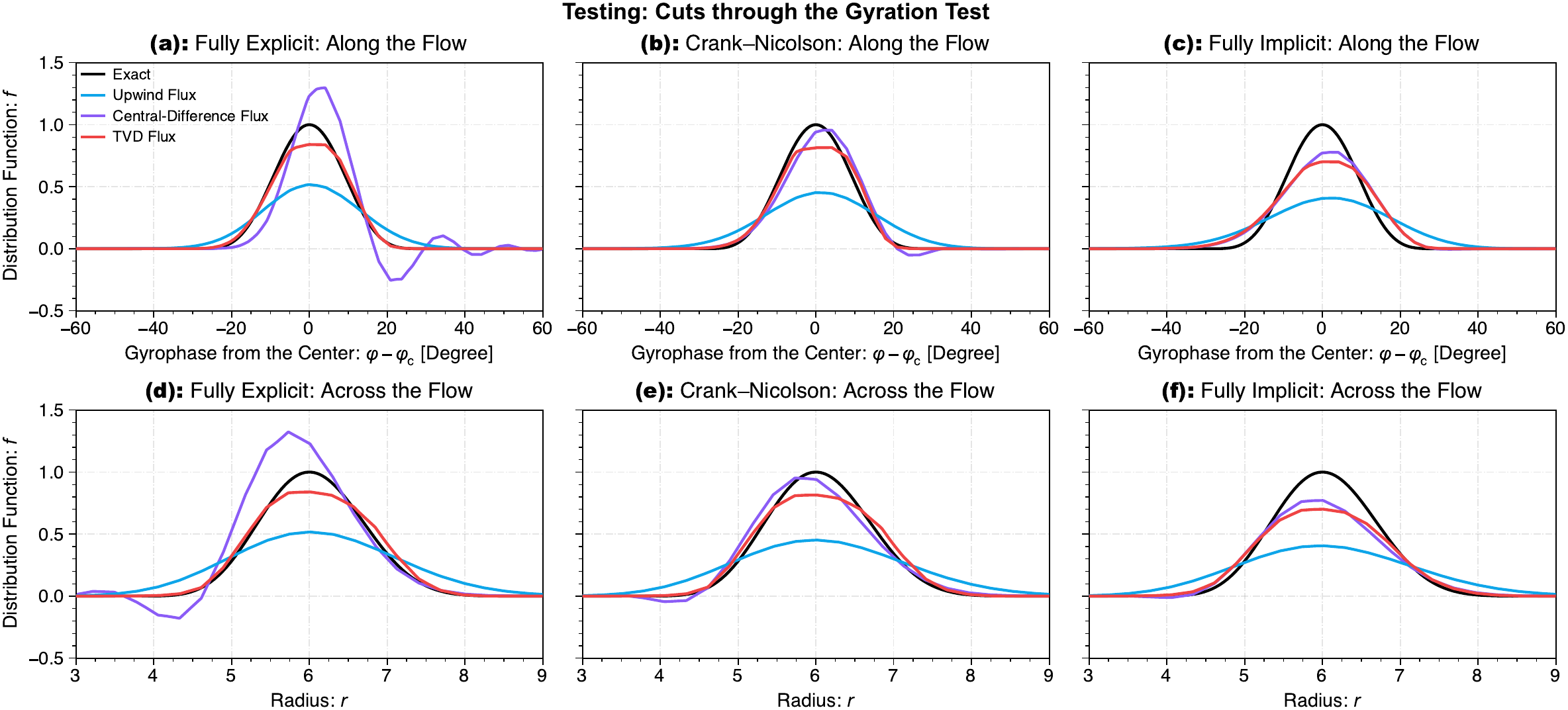}
    \caption{Distribution function profiles in the relativistic gyration test at $t = 2\pi$ on $100 \times 100$ cells with $C=0.5$. Columns correspond to the fully explicit (forward Euler), Crank--Nicolson, and fully implicit (backward Euler) schemes, respectively. (a)--(c) Azimuthal cuts along the circular trajectory at the distance of $r = \sqrt{q^2+p^2} = 6$. (d)--(f) Radial cuts through the cap center $\varphi_\mathrm{c}$ in the exact solution. Each panel compares the exact solution (black) with the upwind-flux (blue), central-difference-flux (purple), and TVD-flux (red) solutions, as indicated in the legend of panel~(a).} \label{fig3:cuts}
    \end{figure*}

    Figure~\ref{fig3:cuts} compares the upwind-flux, central-difference, and TVD-flux solutions at $t = 2\pi$ on $100 \times 100$ cells with $C=0.5$, with cuts along the phase-space flow in panels~(a)--(c) and across it in panels~(d)--(f). Numerical diffusion substantially smears the upwind-flux solution, and the central-difference solution oscillates and can become negative. The TVD-flux solution basically preserves the cap profile without spurious oscillations and stays within $0 \leqslant f \leqslant 1$ for all three time-stepping methods.
    Besides, in all these runs, the total particle number $\sum_i N_i$ remains constant to within a relative change below $10^{-10}$.

    \begin{figure*}[tp!]
    \centering
    \includegraphics[width=0.965\hsize]{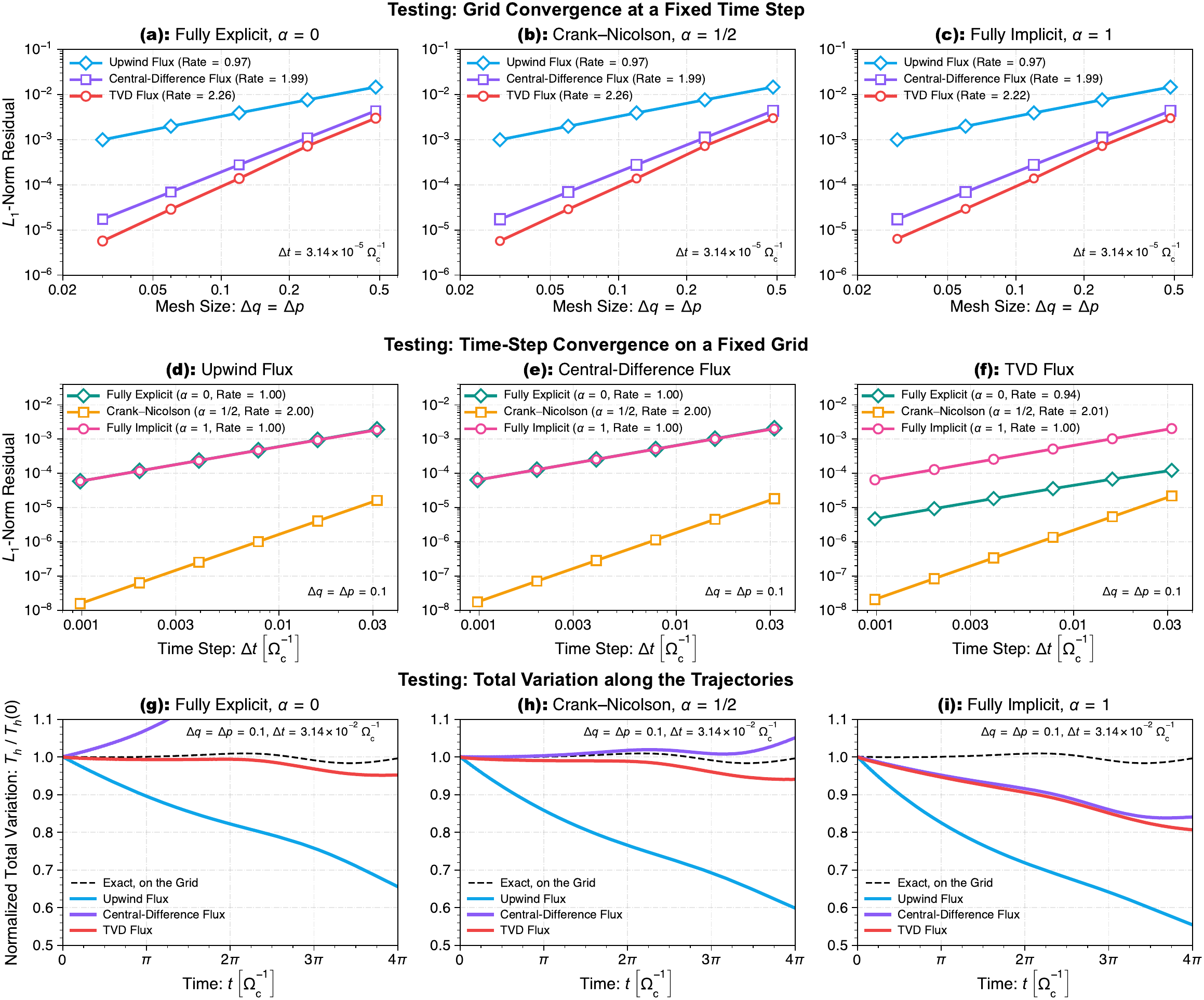}
    \caption{Convergence and total variation in the relativistic gyration test. 
    (a)--(c) Spatial convergence of the $L_1$-norm residual with respect to the exact solution at $t=\pi/8$, with varying mesh spacings $\Delta q = \Delta p$ and fixed $\dt = 3.14\times10^{-5}$. 
    (d)--(f) Temporal convergence of the $L_1$-norm residual with respect to the exact solution at $t=\pi/4$, for varying $\dt$ on a fixed grid with $\Delta q = \Delta p=0.1$. Legends show the fitted convergence rates. 
    (g)--(i) Total variation along the Hamiltonian trajectories, normalized to its initial value, for $\Delta q = \Delta p = 0.1$ and $\dt = 3.14\times10^{-2}$, with dashed curves for the exact solution sampled on the grid.} \label{fig4:conv}
    \end{figure*}

    We examine the spatial and temporal convergence and the TVD property. As shown in Figure~\ref{fig4:conv}(a)--(c), with fixed, sufficiently small time steps, spatial refinement gives $L_1$ convergence rates of $0.97$, $1.99$, and $2.2$--$2.3$ for the upwind-flux, central-difference, and TVD-flux schemes, respectively. In time, the fully explicit (forward Euler) and implicit (backward Euler) methods converge at first order and the Crank--Nicolson scheme at second order (panels (d)--(f)). 
    As anticipated, the total variation calculated along the Hamiltonian trajectory \citep[$T_h$; see also][]{sokolov2023high} of the TVD solution stays below its initial value throughout the test but grows for the central-difference solution with the fully explicit and Crank--Nicolson schemes (panels (g)--(i)). The monotone upwind flux is also TVD, but its strong numerical diffusion causes a large reduction in the total variation. 

    \subsection{Analytical Shock Acceleration} \label{app:shock}

    We consider DSA at a steady, planar shock with a density compression ratio of $4$, which gives an analytical particle spectrum $f = 4(p_0/p)^4$ above the injection momentum $p_0$ \citep{krymskii1977regular, axford1977acceleration, bell1978acceleration1, blandford1978particle, drury1983introduction}.
    Following the test in \citet{sokolov2023high}, we consider a shock propagating at unit speed along a uniform magnetic field into stationary plasma. We hold the grid fixed in the \textit{boosted frame} of reference, comoving with the shock, and solve Equation~\eqref{eqn:parker} for the steady-state particle distribution. 
    In the shock rest frame, the normalized plasma speed and density are $u=-1$ and $\rho=1$ upstream and $u=-1/4$ and $\rho=4$ downstream, with $B=1$ and $\kappa_\parallel=20$, giving a constant $\rho u/B=-1$. We use a logarithmically spaced momentum grid with $100$ bins per decade, and seed particles enter through the upstream boundary in the first bin above $p_0$. Since the shock accelerates particles only toward higher momenta, the implicit linear system in Equation~\eqref{eqn:monolithic} is solved with a single block Gauss--Seidel sweep in each iteration. 

    \begin{figure*}[tp!]
    \centering
    \includegraphics[width=0.96\hsize]{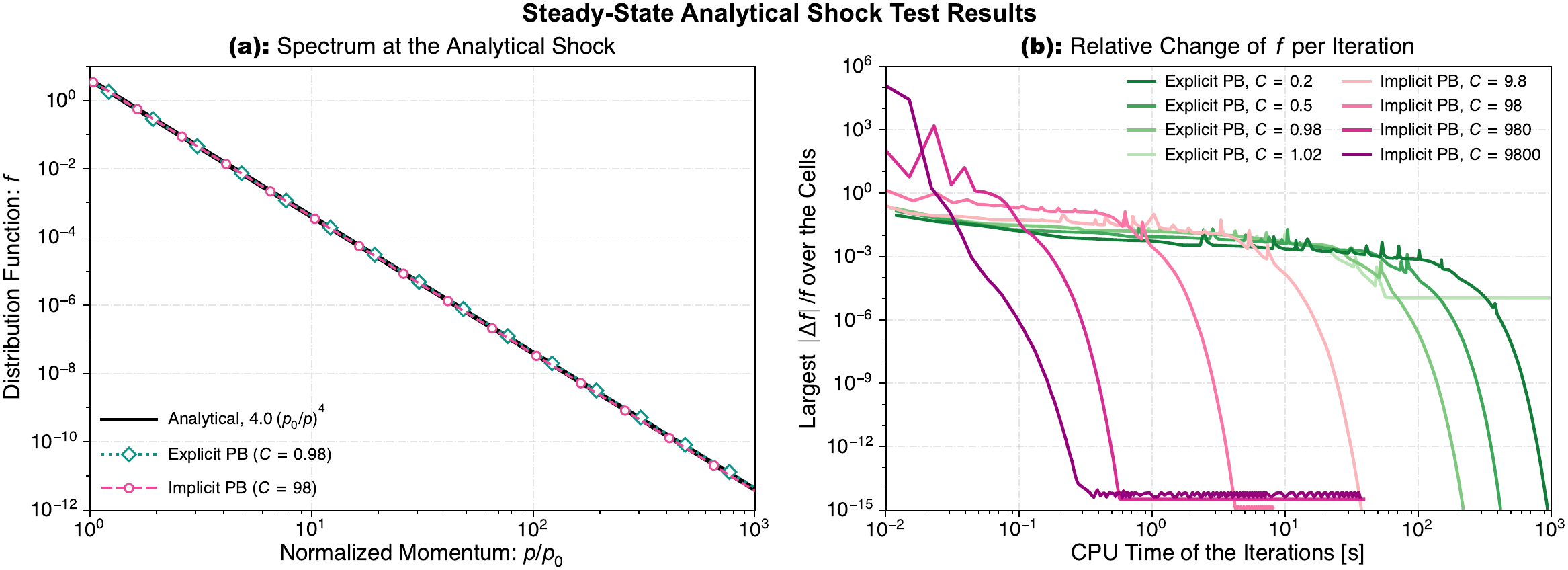}
    \caption{Diffusive shock acceleration spectrum at an analytical shock and convergence toward the steady state. 
    (a) Particle spectra at the shock front obtained with the explicit ($C = 0.98$) and implicit ($C = 98$) PB schemes, compared with the analytical solution, $f = 4(p_0/p)^4$. 
    (b) Largest relative change in $f$ across all cells between successive iterations as a function of CPU time, for the explicit and implicit PB schemes with different CFL numbers indicated in the legend.} \label{fig5:shock}
    \end{figure*}

    Figure~\ref{fig5:shock}(a) presents the particle spectra at the shock front using the implicit and explicit PB schemes, at $C=98$ and $0.98$, respectively, and they agree closely and reproduce the analytical spectral index of $-4$.
    We further conduct convergence tests at different CFL numbers to assess the stability and computational efficiency of the implicit PB scheme at large time steps. For each explicit and implicit run, we monitor the largest relative change in $f$ between successive iterations and record the cumulative CPU time, as displayed in Figure~\ref{fig5:shock}(b). 
    For example, to reach a tolerance of $10^{-12}$, the explicit scheme requires $\sim\! 200~\mathrm{s}$ ($>10^4$ iterations) at $C=0.98$, and the run at $C=1.02$ fails to converge. By contrast, the implicit scheme requires only $\sim\! 4~\mathrm{s}$ ($<\! 500$ iterations) at $C=98$ and $<\! 0.3~\mathrm{s}$ ($\sim\!30$ iterations) at $C=9800$, giving CPU speedups of $\sim\! 50$ and $>\! 700$, respectively, and demonstrating its efficiency in solving steady-state problems.

\bibliographystyle{aasjournalv7.1}
\bibliography{ref_paper}

\end{document}